\documentclass[11pt]{article}
\usepackage[dvips]{graphicx}\usepackage{latexsym}
\usepackage[]{amsmath}\usepackage{amsfonts,amssymb}
\usepackage{color}

\newtheorem{theorem}{Theorem}\newtheorem{lemma}{Lemma}
\newtheorem{definition}{Definition}\newtheorem{proposition}{Proposition}

\def\squarebox#1{\hbox to #1{\hfill\vbox to #1{\vfill}}}
\def\qed{\hspace*{\fill}        \vbox{\hrule\hbox{\vrule\squarebox{.667em}\vrule}\hrule}\smallskip}

 \newcommand{\bs}{\bigskip} 
 \newcommand{\n}{\noindent} 
 \newcommand{\hs}[1]{\hspace*{ #1 mm}}

\newcommand{\ignore}[1]{}

\begin{document}
\pagestyle{plain}
\begin{center}
{\Large {\bf Reconstructing Strings from Substrings with Quantum Queries}}
\bs\\

{\sc Richard Cleve}$^1$ \hspace{5mm}
{\sc Kazuo Iwama}$^2$ \hspace{5mm} 
{\sc Fran\c{c}ois Le Gall}$^3$ \hspace{5mm} 
{\sc Harumichi Nishimura}$^4$ \hspace{5mm}\\
{\sc Seiichiro Tani}$^5$ \hspace{5mm} 
{\sc Junichi Teruyama}$^2$ \hspace{5mm} 
{\sc Shigeru Yamashita}$^6$ 

\

{\small
$^1${Institute for Quantum Computing and School of Computer Science, University of Waterloo, Canada; \\
Perimeter Institute for Theoretical Physics, Canada}; \\
{\tt cleve@cs.uwaterloo.ca} 

$^2${School of Informatics, Kyoto University, Japan}; \\
{\tt $\{{\tt iwama,teruyama}\}$@kuis.kyoto-u.ac.jp} 

$^3${Department of Computer Science, The University of Tokyo, Japan}; \\
{\tt legall@is.s.u-tokyo.ac.jp}

$^4${School of Information Science, Nagoya University, Japan}; \\
{\tt hnishimura@is.nagoya-u.ac.jp}

$^5${NTT Communication Science Laboratories, NTT Corporation, Atsugi, Japan}; \\
{\tt tani.seiichiro@lab.ntt.co.jp} 

$^6${College of Information Science and Engineering, Ritsumeikan University, Japan}; \\
{\tt ger@cs.ritsumei.ac.jp} 

}
\end{center}
\bs

\n{\bf Abstract.}\hs{1} 
This paper investigates the number of quantum queries made
to solve the problem of reconstructing an unknown string from its  
substrings in a certain query model. More concretely, the goal of the problem is
to identify an unknown string $S$ by making queries of the following  
form: ``Is $s$ a substring of $S$?'', where $s$ is a query string over the given alphabet.
The number of queries required to identify the string $S$ is the query complexity of this problem.

First we show a quantum algorithm that exactly identifies the string $S$
with at most $\frac{3}{4}N + o(N)$ queries, where $N$ is the length of $S$.
This contrasts sharply with the classical query complexity~$N$.
Our algorithm uses Skiena and Sundaram's classical algorithm 
and the Grover search as subroutines.
To make them effectively work, we develop another subroutine 
that finds a string appearing only once in $S$, which may have an  
independent interest. 
We also prove two lower bounds.  The first one 
is a general lower bound of $\Omega(\frac{N}{\log^2{N}})$,
which means we cannot achieve a query complexity of 
$O(N^{1-\epsilon})$ for any constant $\epsilon$.  
The other one claims that if we cannot use queries of length 
roughly between $\log N$ and $3 \log N$, then we cannot achieve a query
complexity of any sublinear function in $N$.

\section{Introduction}
For an input of length $N$, we usually assume that the time complexity
of any algorithm ${\cal A}$ is at least $N$, since ${\cal A}$ needs
$N$ steps only to read
the input.  However, especially recently, there have been increasing
demands for studying algorithms that run in significantly less than $N$
steps by sacrificing the exactness of the computation.  In this case,
we obviously need some mechanism for algorithms to obtain the input,
since it is no longer possible to read all the input bits sequentially. 
{\it Oracles} are a popular model for this purpose.
The most standard oracle is so-called an {\it index oracle}, a mapping $f$
from $\{0,1, \ldots, N-1\}$ into $\{0,1\}$ such that $f(i)$ returns the $i$th bit of the input. 
Thus, we need $N$ oracle calls in order to get all the input bits.  
A little surprisingly, however, some Boolean functions can be computed, 
with high success probability, using oracle calls much less than $N$ times.
For example, a balanced AND-OR tree can be computed with $O(N^{0.753\ldots})$ oracle calls 
with high success probability~\cite{Sni85}.

This interesting fact becomes even more impressive if we are allowed
to use {\it quantum} oracles.  Due to the famous Grover
search~\cite{Gro96}, we need only $O(\sqrt{N})$ oracle calls to compute the Boolean-OR
function with high success probability, or a quadratic speed-up against its classical version 
(classically we need $\Omega(N)$ calls). This result is widely known as one of the two most remarkable examples
claiming the superiority of quantum computation over classical computation (the
other is Shor's integer factorization algorithm~\cite{Shor94}).

To compute the Boolean-OR, it suffices to find at least one true value in
the input bits.  The {\it oracle identification problem}, or the {\it
string reconstruction problem}, is more general and
more difficult, namely it requires us to recover all the $N$ bits of
the input (thus any Boolean function can be computed without any
additional oracle calls). The quantum index oracle is still nontrivially
powerful for this problem; Ref.~\cite{D98} shows that 
$N/2 + O(\sqrt{N})$ oracle calls are enough for this problem, while we obviously need $N$ queries in the
classical counterpart.
There are different types of oracles that are much more powerful for
this most general problem.  The quantum {\it IP oracle}~\cite{BV97}, a
function $g$ from $\{0,1\}^N$ into $\{0,1\}$ such that $g(q)=q\cdot x$ for the
input string $x$, needs only one oracle call to recover $x$ while its classical
counterpart $N$ oracle calls.  Recently, Ref.~\cite{INRT10} studied the
{\it balance oracle}, which models the balance scale to be used 
for the counterfeit coin problem
(i.e., for finding the $k$ counterfeit coins in $N$ coins),
and shows its quantum version can be solved with $O(k^{1/4})$ oracle calls 
while the classical version requires $\Omega(k \log(N/k))$ calls,
where $k$ is the number of 1's in $x$.

In 1993, Skiena and Sundaram~\cite{SS93} showed that 
$N+\Theta(\sqrt{N})$ (classical) queries are sufficient to reconstruct the
hidden string $x$ if we use a {\it substring oracle} or an {\it S-oracle}, in short.
This oracle, $h(q)$, which returns 1 if the query string $q$ is a substring of $x$,
and 0 otherwise, had been quite popular in the algorithm community.
For example it plays an important role in computational biology such
as sequencing by hybridization~\cite{DC87,LFKKSM88,PL94}.  One should
notice that there is no obvious way (even regardless of its
efficiency) of using this oracle for string reconstruction ($h(q)$
probably returns yes almost always if $|q|$ (the length of $q$) is short, say two or
three, and no almost always if $|q|$ is, say, 10).  Thus Skiena and
Sundaram's result was highly appreciated, whose basic idea is as
follows: Suppose that we already know that a substring $s$ exists in
the input $x$.  Then we ask the oracle if $s1$ is a substring. If the
answer is yes, we can increase the length of a confirmed substring by
one.  Otherwise, we know $s0$ is a substring or $s$ is at the right
end of $x$. Just assume the former and check the latter occasionally
and we can get the above bound.  It is almost tight information-theoretically.

Now here is our question in this paper: Is quantum also more powerful 
than classical computation for this oracle, and how much is it if yes?  
One might say the answer is easy: Instead of asking if $s1$ is a substring, 
we ask which of $x00$, $x01$, $x10$ and $x11$ is a substring 
using the $1/4$-Grover search~\cite{BBHT98}. 
Since $1/4$-Grover needs just one query, we can increase 
the confirmed substring by two per call, 
or we would get a roughly $N/2$ upper bound.  
Unfortunately it immediately turns out that this does not work, 
since more than one of the four candidates may be (correct) substrings of $x$ 
at the same time (recall that $1/4$-Grover only works for a unique solution).

{\bf Our Contribution.}
Here is our main result in this paper:
\begin{theorem}\label{thm:upper}
The quantum query complexity for identifying S-oracle is 
at most $\frac{3}{4}N + O(\sqrt{N} \log N)$.
\end{theorem}

Therefore, the quantum algorithm is better than its classical counterpart by a factor of $3/4$.  
Notice that our algorithm is {\it exact} as well as the classical one in~\cite{SS93}.
To cope with the difficulty mentioned above, 
we use Skiena and Sundaram's algorithm until the
confirmed substring gets to a certain length, then change our 
algorithm to the one based on $1/4$-Grover.  There still exists the
possibility of multiple solutions, say $s00$ and $s01$, but now we can
assume that $s$ is pretty long or those two strings need to overlap if
they are both solutions. This gives us a lot of information about the string $s$, 
which basically changes the problem into a certain kind of
string manipulation problem that has a long history in theoretical computer science.
By using this information,
we construct the procedure which makes the situation that $1/4$-Grover is useful.

Our strong conjecture is that our problem needs at least a linear number of queries.  
Our basic idea is to use the quantum adversary method~\cite{Amb02,Zha04},
but it turns out that the fact that there is a wide range (one to $N$) 
in the length of query strings makes its direct application hard. 
We bypass this difficulty with two different approaches: 
The first one is to introduce a new query model, an {\it anchored substring} oracle, 
which is something between our substring oracle and 
the standard index oracle and makes it possible to 
exploit the basic ideas of the adversary method for the latter. 
This gives us the following theorem. 
See Appendix~\ref{app:low} for the proof.
\begin{theorem}\label{thm:lower}
The quantum query complexity for identifying an S-oracle is $\Omega\left(\frac{N}{\log^2 N} \right)$.
\end{theorem}
This theorem means that there are no algorithms with a query complexity of
$N^{1-\epsilon}$ for any positive constant $\epsilon$.
The second one is to prohibit a small range of length for available queries.
\begin{theorem}\label{thm:res}
Suppose that we cannot use queries of length
$\log N - 1 - 2\log\log N$ to $3\log N$.
Then the problem of identifying an S-oracle needs $\Omega(N)$ queries.
\end{theorem}
This theorem says that we need to use
queries of the range of length 
between $\log N - 1 - 2\log\log N$ and $ 3 \log N$
``effectively'' to achieve a sublinear bound. 
See Appendix~\ref{app:res} for the details.

{\bf Related Work.}
There have been many studies achieving quantum linear speedups.
As mentioned already, a most celebrated one is due to van Dam~\cite{D98},
who presented a quantum algorithm for identifying the oracle
by $\frac{N}{2} + O(\sqrt{N})$ queries. This is optimal up to a
constant factor since the lower bound $\frac{N}{4}$ was obtained
by Ambainis \cite{Amb99}. Another example is ordered search,
that is, to find a target in a sorted list of $N$ items.
Farhi et al.~\cite{FGGS99} invented a quantum algorithm that makes
at most $c\log N$ queries with $c\approx 0.53$
(note that any classical algorithm needs at least $\log N$ queries),
and the constant $c$ was subsequently improved \cite{CLP07,BH07}.
These linear speedups were also turned out to be tight 
(up to a constant factor) by the lower bound results in \cite{Amb99-2,HNS02,CL08} 
which improved the previous lower bounds of \cite{BW99,FGGS98}.

There are no quantum studies based on substring oracles, and few ones about 
string manipulation previously. One of them is a quantum algorithm given
by Ramesh and Vinay \cite{RV03} which determines
if a given pattern appears in a given text by combining Grover's search
with a classical string matching technique called deterministic sampling.

\section{Upper Bounds}\label{sec:upper}

Now we give the definition of our oracle model.
We call it a {\it substring oracle}, or simply an {\it S-oracle}.

\begin{definition}
A {\it substring oracle}, or an {\it S-oracle}, in short, 
is a binary string $x=x_0 \cdots x_{N-1} \in \{0,1\}^N$.
A query to an S-oracle is given as a string $s \in \bigcup_{k=1}^{N}\{0,1\}^{k}$.
The answer from the S-oracle is a binary value $\chi(x;s)$ 
defined as follows: If $x$ has $s$ as substring, that is, 
there exists an integer $i$ such that $x_{i+k-1}=s_{k}$ for all $1 \leq k \leq |s|$
then $\chi(x;s)=1$ and otherwise $\chi(x;s)=0$.
In the quantum computation an S-oracle is viewed as the unitary transformation $O_{S,x}$
that transforms $|s\rangle |a\rangle$ to $|s\rangle |a \oplus \chi(x;s)\rangle$.
\end{definition}

To give the proof of Theorem~\ref{thm:upper}, we define some notations on strings.
The string representing the concatenation of strings $u$ and $v$ will be denoted $uv$.
When $z=uv$, we call $u$ a {\em prefix} of $z$ and call $v$ a {\em suffix} of $z$. 
A string $v$ is called a {\em presuffix} of a string $w$ if $v$ is a prefix of $w$ and also a suffix of $w$.
The string formed by concatenating $i$ copies of $z$ will be denoted $z^i$. 
A string $t$ is called the {\it periodic string} of a string $a$ 
if $t$ is the shortest string such that $a_i = t_{(i \mod{|t|})}$ for all $i$
(or, equivalently, $t$ is the shortest string such that $a$ can be written as $a = t^kb$ for some integer $k$ and some prefix $b$ of $t$).

\subsection{Basic Ideas and Algorithms}

Before the full description of our algorithm, we present the basic idea. 
The algorithm has three main steps. At the first step, we use Skiena and Sundaram's algorithm~\cite{SS93}, 
which extends a substring in the oracle string $x$ by one letter with one query.  
At the second step, we extend the substring $z$ obtained by the first step 
to a string $z_{out}$ so that $z_{out}$ can appear only once in $x$. 
Note that the first and second steps are implemented classically. 
The third step is quantum: we apply Grover's search algorithm~\cite{Gro96} 
under the special case that is called {\it $1/4$-Grover search}~\cite{BBHT98}. 
Recall that the $1/4$-Grover search can find a solution surely with only one query
in the case when we know there is only one solution out of four candidates.
Since the second step assures that the substring $z_{out}$ appears only once in $x$, 
there is exactly one substring of $x$ in $\{ 00z, 01z, 10z, 11z \}$ for any string $z$ 
that extends $z_{out}$ unless $z$ corresponds to the leftmost part of $x$.
So the $1/4$-Grover search can extend the substring by two letters with only one query. 
If we know that $z$ is a prefix of the oracle string, we run the $1/4$-Grover search 
for $\{z00, z01, z10, z11 \}$. 
 
The second step is the most technical and it is also essential to implement 
the third step successfully. A key idea for obtaining the substring 
appearing only once is relatively simple; extending $z$ by its periodic string. 
For instance, we assume $x=10 101 101 101 101 1 1110$. 
Then the substring $z= 101 101 1$ appears three times in $x$. 
The periodic string $t$ of $z$ is $t=101$. Let us extend $z$ by $t$ 
as long as possible such that $t^iz$ is still a substring of $x$.
In the example we get a substring $t^2z = 101 101 101 101 1$, which appears only once in $x$. 
Now the difficulty is to make the string $z$ obtained by the first step 
as short as possible, which improves the complexity of the algorithm. 
Another key idea for this difficulty is to analyze what happens 
when $z$ appears twice in $x$.
When a substring $z$ with length $>N/2$ appears twice in $x$, 
these occurrences of $z$  must be partially overlapping, and $x$ has a substring $uvw$ 
such that $z=uv=vw$. A key property is that the overlapped string $v$ 
is a presuffix of $z$. Using these key ideas we can construct 
the algorithm by starting from the substring $z$ of length $>N/2$.

Now we give an exact algorithm {\it Identify} and its subroutine {\it MakeOnce}.

\noindent{\bf Algorithm {\it Identify}}\\ 
Input : an S-oracle $O_{S,x}$. \\
Output : the oracle string $x$.\\ 
\: Step 1. Find a substring $z$ of length $\lceil N/2\rceil+1$ using Skiena and Sundaram's algorithm~\cite{SS93}.\\ 
\: Step 2. Run the algorithm {\it MakeOnce} on input $z$. Let $z_{out}$ be the output.\\ 
\: Step 3. Repeat extending $z_{out}$ to the left by $2$ letters using the $1/4$-Grover search.
   Check whether the extended string is a substring of $x$ after every $\sqrt{N}$ applications 
   of the $1/4$-Grover search. If not, we know that a prefix of $x$ is obtained 
   between the current check point and the previous check point. 
   Then, find this prefix by binary search.\\
\: Step 4. Repeat extending the current substring to the right by $2$ letters using the $1/4$-Grover search, 
and stop when the length of the substring becomes $N-1$ or $N$. 
If the length is $N-1$, use a classical query to find the last bit.
\\(End of Algorithm {\it Identify})

\noindent {\bf Algorithm {\it MakeOnce}}\\
Input  : a string $z$ (a substring of $x$, $|z| > N/2$); an S-oracle $O_{S,x}$.\\
Output : a substring $z_{out}$ that appears only once in $x$.\\ 
\: Step 1. $T_0:=\emptyset$. $A_0:=\emptyset$. $l:=1$. $z_1:=z$. ($A_l$ is used for the analysis.)\\
\: Step 2. Repeat Steps 2.1--2.7.\\
\:\:\: Step 2.1. Find the shortest string $a_l$ satisfying the following conditions.\\
\:\:\: \quad (i)  $a_l$ is a presuffix of $z_l$.\\
\:\:\: \quad (ii) The periodic string of $a_l$ is not in $T_{l-1}$.\\
\:\:\: \quad If there is no such string, go to Step 3. Let $t_l$ be the periodic string of $a_l$.
$T_l := T_{l-1} \cup \{t_l\}$. $A_{l} := A_{l-1} \cup \{a_l\}$.
\\
\:\:\: Step 2.2. Find the largest integer $i$ such that $(t_l)^i z_l$ is also a substring of $x$. 
Define $z'_l := (t_l)^i z_l$.
\\
\:\:\: Step 2.3. Let $j$ be the largest integer such that $z'_l=u t_l^ja_l$ for some string $u$.
\\
\:\:\: Step 2.4. Let $h$ be the largest integer such that $z'_l = t_l^h a_l w$ for some string $w$.
\\
\:\:\: Step 2.5. If $z'_{l}=t_l^j a_l$ or $h < j$, then $z_{l+1}:=z'_l$ and go to Step 2.7.
\\
\:\:\: Step 2.6. Find the largest integer $k$ such that $u^k z'_l$ is also a substring of $x$. 
Define $z_{l+1} := u^k z'_l$.
\\
\:\:\: Step 2.7. $l:=l+1$.
\\
\: Step 3. $l_{max}:=l$. $z_{out}:=z_{l_{max}}$. $T:=T_{l_{max}-1}$. $A:=A_{l_{max}-1}$.  
($l_{max}$, $T$ and $A$ are used for the analysis.)
\\(End of Algorithm {\it MakeOnce})

\subsection{Analysis of {\it MakeOnce}}

In this section, we give the analysis of {\it MakeOnce}. First, a number of properties are given for the analysis.
See Appendix~\ref{app:lem} for the proof.

\begin{lemma}\label{lem:pro}
For any $l < l_{max}$, {\it MakeOnce} satisfies the following properties.
\begin{enumerate}
\item\label{lem:pro1} $a_{l} \in A$ is represented as $a_l = t_l b_l$, where $t_l \in T$ and $|b_l|<|t_l|$.
\item\label{lem:pro2} $z'_{l}$ and $a_{l} \in A$ are prefixes of $z_{l+1}$.
\item\label{lem:pro3} $z_{l}$ (and hence $a_{l} \in A$) is a suffix of $z_{l+1}$.
\item\label{lem:pro4} $a_l$ is a presuffix of $a_{l+1}$ and $|a_{l+1}| > |a_l|$.
\item\label{lem:pro5} $|a_{l+1}| \geq |a_l| + |t_l|$.
\item\label{lem:pro6} $|t_{l+1}| > |t_{l}|$.
\item\label{lem:pro7} At step 2.6, $|u| > |t_{l}|$.
\item\label{lem:pro8} $l_{max}=O(\sqrt{N})$.
\end{enumerate}
\end{lemma}

Now we analyze the query complexity and the correctness of {\it MakeOnce}.  
In what follows, we refer to the properties $\ref{lem:pro1}$--$\ref{lem:pro8}$ of Lemma~\ref{lem:pro}
as simply the properties $\ref{lem:pro1}$--$\ref{lem:pro8}$.

\begin{proposition}\label{querycomplexity_makeonce}
{\it MakeOnce} uses at most $O(\sqrt{N}\log{N})$ queries.
\end{proposition}

{\it Proof.} 
To obtain $z_{out}$, we need queries only at Step 2.2 and Step 2.6.
These steps can be implemented by binary search to find $t_l^iz_l$ and $u^kz'_l$, 
which use $O(\log N)$ queries. Since the number of repetitions of Step 2 is $O(\sqrt{N})$ 
by the property~\ref{lem:pro8}, the total number of queries is $O(\sqrt{N}\log{N})$.
\qed

\begin{proposition}\label{correctness_makeonce}
The output $z_{out}$ of {\it MakeOnce} appears exactly once in $x$.
\end{proposition}

Proposition~\ref{correctness_makeonce} is proved by contradiction. 
We assume that $z_{out}$ appears twice in $x$. 
Since $|z_{out}|>\frac{N}{2}$, $x$ has a substring $uvw$ such that 
$z_{out} = uv = vw$, where $|u|=|w|>0$. Then we can see that $v$ has the following special form.

\begin{lemma}\label{cla:1}
$v=t_l^m a_l$ for some $l>0$ and $m \geq 0$ where $t_l \in T$ and $a_l \in A$.
\end{lemma}

{\it Proof.}
First we should notice that $z_{out}$ has no substring which satisfies 
the conditions at Step 2.1
since we go to Step 3 and $z_{out}$ is output only when there is no string satisfying 
the conditions at Step 2.1.
On the contrary, $v$ is a presuffix of $z_{out}$, 
which means that $v$ satisfies the condition (i) of Step 2.1.
This implies that $v$ does not satisfy the condition (ii) of Step 2.1.
That is, the periodic string of $v$ must be $t_l$ in $T$ for some $l$.
Hence, it is represented as $v = t_l^{m'} y$ where $l>0$, $m' > 0$, 
$t_l \in T$, and $|y| < |t_l|$.

For $a_l \in A$, let $b_{l}$ be the string such that $a_l = t_l b_l$ 
and $|b_l| < |t_l|$ as guaranteed by the property~\ref{lem:pro1}.
By the property~\ref{lem:pro3}, $a_l$ is a suffix of $z_{out}$.
Also, $v$ is a suffix of $z_{out}$. Thus $y$ has suffix $b_l$ or $b_l$ has suffix $y$. 
Now we show that $y=b_l$ by contradiction. 
Assuming $|y|<|b_l|$, it must hold that $t_l b_l = y' t_l y$ for some $y'$ such that $|y'| < |t_l|$.
Then the length of the periodic string of $a_l$ is at most $|y'|$, 
which contradicts that $t_l$ is the periodic string of $a_l$.
Assuming $|y|>|b_l|$, $y' t_l b_l = t_l y$ for some $y'$ such that $|y'| < |t_l|$.
Then the length of the periodic string of $a_l$ is at most $|y'|$,
which also leads to a contradiction.

By the above arguments, $v$ is represented as $v = t_l^{m'} b_l = t_l^{m} a_l$ where $m = m'-1$.
\qed

The main statement for the correctness of {\it MakeOnce} is now stated as follows.
(In the rest of this section, we assume that $u$, $w$, $u'$ and $w'$ have positive length.)

\begin{lemma}\label{lem:6}
For any $l \leq l_{max}$, any $c<l$ and $m \geq 0$, $x$ has no substring $u t_c^m a_c w$
such that $z_l = u t_c^m a_c = t_c^m a_c w$, $t_c \in T_{l-1}$ and $a_c \in A_{l-1}$.
\end{lemma}

Then, by the assumption that $z_{out}$ ($=z_{l_{max}}$) appears twice in $x$, 
Lemma~\ref{cla:1} implies that $x$ has a substring $ut_l^m a_lw$ 
for some $0< l \leq l_{max}-1$ and $m\geq 0$, which contradicts Lemma~\ref{lem:6}. 
This completes the proof of Proposition~\ref{correctness_makeonce}.

What remains is the proof of Lemma~\ref{lem:6}. We prove the statement by induction on $l$. 
The case of $l=1$ is easy. In this case, $T_0=A_0=\emptyset$, $z_1 = u = w$ and 
$|z_1|=|z| > \frac{N}{2}$. Hence $x$ does not have a substring $uw=z_1z_1$. 
Next we assume that the statement holds for $l$, and show that the statement holds for $l+1$. 
For this purpose, we first show the following lemma:
\begin{lemma}\label{cla:3}
If $x$ has $u' t_{c'}^{m'} a_{c'} w'$ as a substring such that 
$z_{l+1} = u't_{c'}^{m'} a_{c'} = t_{c'}^{m'} a_{c'} w'$ for some $c' < l$ and $m' \geq 0$, 
then $x$ also has $u t_c^m a_c w$ such that $z_{l} = u t_c^m a_c = t_c^m a_c w$ for some $c \leq c'$ and $m \geq 0$.
\end{lemma}

{\it Proof.} 
By the property~\ref{lem:pro3}, $z_{l}$ is a suffix of $z_{l+1}$.
By the assumption, $z_{l+1}$ appears twice in $x$, and hence $z_{l}$ also appears twice in $x$.
Since $|z_l|>N/2$, $x$ has a substring $uvw$ with $z_l=uv=vw$.
Let $t$ be the periodic string of $v$.
Then $v$ is represented as $t^{m_v}b$ for some $m_v>0$, 
where $|b|<|t|$ and $b$ is a prefix of $t$. 
Note that $|t|\leq |t_{c'}|$ since $v$ is a suffix of $t_{c'}^{m'}a_{c'}$.

Now we show that there is $c \leq c'$ such that $t = t_c$ and $b = b_c$, 
where $b_c$ is the string such that $t_c b_c = a_c$ as guaranteed 
by the property~\ref{lem:pro1}.
First we show $t\in T_l$, which means that $t=t_c$ for some $c\leq c'$
by $|t|\leq |t_{c'}|$ and property~\ref{lem:pro6}. 
For contradiction, we assume that $t \notin T_l$ (and hence $\notin T_{l-1}$).
Note that since $v$ satisfies the condition (i) of Step 2.1 for the $l$-th loop 
(i.e., $v$ is a presuffix of $z_l$), $tb$ also satisfies this condition. 
Then, by $t \notin T_{l-1}$, $tb$ satisfies the conditions (i) and (ii) at Step 2.1.
Since $a_l$ is the shortest string satisfying the conditions (i) and (ii) at Step 2.1, $|t b|\geq|a_l|$. 
This means that $a_l$ is a prefix of $tb$.
Then we have $|t_l| \leq |t| \leq |t_{c'}|$ and $c' < l$.
This contradicts the property~\ref{lem:pro6}.
Second we show $b=b_c$. 
To this end, it suffices to show $|b| = |b_c|$ 
because both $b$ and $b_c$ are prefixes of $t=t_c$. 
Assume that $|b| < |b_c|$. Then $y t_c b = t_c b_c$ for some $y$ such that $|y| < |t_c|$
since $tb=t_cb$ is a suffix of $z_l$ and also, 
by property~\ref{lem:pro3}, $a_c=t_cb_c$ is a suffix of $z_l$.
Then, the length of a periodic string of $a_c$ is at most $|y|$,
which contradicts that $t_c$ is a periodic string of $a_c$.
By a similar argument, we also have a contradiction assuming that $|b| > |b_c|$. 
Thus $|b| = |b_c|$.

We conclude that $t b = t_c b_c = a_c$, which completes the proof of Lemma~\ref{cla:3}.
\qed

Lemma~\ref{cla:3} and the induction hypothesis imply: 
For any $c<l$ and $m \geq 0$, $x$ has no substring $u t_c^m a_c w$ such that 
$z_{l+1} = u t_c^m a_c = t_c^m a_c w$, $t_c \in T_{l}$ and $a_c \in A_{l}$.
We now show another lemma.
\begin{lemma}\label{cla:for_induction_2}
For any $m \geq 0$, $x$ has no substring $u' t_l^m a_l w'$ such that $z_{l+1} = u' t_l^m a_l= t_l^m a_l w'$.
\end{lemma}

{\it Proof.} 
For contradiction, we assume that there is an $m\geq 0$ such that 
$x$ has a substring $u' t_l^m a_l w'$ satisfying $z_{l+1} = u' t_l^m a_l= t_l^m a_l w'$. 
Then we lead to a contradiction for all the possible three cases at Step 2.5:
(1) $z'_l=t_l^ja_l$; (2) $h<j$; (3) the other case. 

In case (1), since $t_l^j a_l=z_{l+1}= u' t_l^m a_l$, we have $m<j$ and $u'=t_l^{j-m}$. 
Then $u' t_l^m a_l w' = t_l^{j-m} z'_l$ is a substring of $x$, 
which contradicts the maximality of $i$ for $z'_l = t_l^i z_l$ at Step 2.2.

In case (2), $h<j$ and $z_{l+1}:=z'_l=ut_l^ja_l$ for some $u$. 
Note that $m\leq h$ since $h$ is taken as the largest integer 
such that $z'_l = t_l^{h} a_l w$ for some $w$ at Step 2.4. 
Thus $j>m$ and hence $u' t_l^{m} a_l w' = u t_l^{j-m} z'_l$.
This implies that $ut_l^{j-m}z'_l$ and hence 
$t_l^{j-m}z'_l$ are included in $x$, which contradicts
the maximality of $i$ for $z'_l = t_l^i z_l$ at Step 2.2.

In case (3) where $h\geq j$, we take the largest integer $k$ 
such that $u^k z'_l = u^{k+1} t_l^j a_l$ is a substring of $x$, 
and let $z_{l+1} := u^{k+1} t_l^j a_l$ at Step 2.6. 
Notice that $u$ does not have suffix $t_l$ 
and $|u| > |t_l|$ by the property~\ref{lem:pro7}.
This implies that if $z_{l+1} = u^{k+1} t_l^j a_l$ has 
a suffix $t_l^{j'} a_l$ then $j' \leq j$. 
By the assumption, $z_{l+1}$ has $t_l^m a_l$ as a suffix, 
which means $m\leq j$. Moreover, we can show that $m=j$:
Since $z'_l$ is a prefix of $z_{l+1}$ by the property~\ref{lem:pro2}, 
there is a string $w''$ such that $z_{l+1} = z'_l w''$. 
Then $x$ includes $u' t_l^m a_l w' = u^{k+1} t_l^{j-m} z'_l w''$.
However, Step 2.2 means that $x$ does not have a $t_lz'_l$, 
which implies that $m=j$. Then $x$ have a substring $(u^{k+1} t_l^{j} a_l) w'
=u^{k+1} (t_l^{m} a_l w')= u^{k+1}z_{l+1}=u^{2k+1}z'_l$, which contradicts 
the maximality of $k$ at Step 2.6.
\qed

By the above two lemmas, it has been shown that for any $c<l+1$ and $m \geq 0$, 
$x$ has no substring $u t_c^m a_c w$ such that $z_{l+1} = u t_c^m a_c = t_c^m a_c w$, 
$t_c \in T_l$ and $a_c \in A_l$. That is, the statement of Lemma~\ref{lem:6} 
for case $l+1$ holds under the assumption that 
it holds for case $l$. Now the proof of Lemma~\ref{lem:6} is completed.

\subsection{Analysis of {\it Identify}}

First, by following the basic idea described in Section~2.1,
the correctness of {\em Identify} is easily verified.
The output $z_{out}$ of {\em MakeOnce} appears in $x$ only once
by Proposition \ref{correctness_makeonce}. This guarantees that
the 1/4-Grover search can extend $z$ by two letters successfully
in Steps 3 and 4 unless the current string reaches the left or right end.
Moreover, the algorithm knows if the string reaches the ends 
by the regular checking in Step 3 or by the current length in Step 4.

Second, we analyze the number of queries used in {\it Identify}.
At Step 1, we find a substring of length $\lceil \frac{N}{2} \rceil +1$
by extending a string by one letter with one query.
Then the number of queries at Step 1 is $\lceil \frac{N}{2} \rceil +1$.
At Step 2, the subroutine {\it MakeOnce} uses $O(\sqrt{N}\log{N})$ queries
by Proposition~\ref{querycomplexity_makeonce}.
At Steps 3 and 4, we extend a substring of length longer than $\frac{N}{2}$
by two letters with one query. Note that the number of checking whether it is a substring of $x$ is $O(\sqrt{N})$.
Thus the number of queries at Steps 3 and 4 is at most $N/4 + O(\sqrt{N})$.
Therefore, the total number of queries is at most $\frac{3N}{4} + O(\sqrt{N}\log N)$. 

Now the proof of Theorem~\ref{thm:upper} is completed. 

\section{Conclusion}
Obvious future works are a (possible) improvement of the constant factor for the upper bound and a challenge to a linear lower bound (we strongly believe there are no sublinear algorithms).  For the former,
one possibility is to exploit a parity computation as was done in~\cite{D98,INRT10}. 
However, we do not have any indication that parity is substantially easier than reconstruction itself for substring oracles. For the latter we at least need to get rid of the reduction of Section~\ref{app:red} since we have already lost a
$\log N$ factor by that.  Different approaches like the polynomial method~\cite{BBC98} do exist as a possibility, but we have no idea on this direction, either, at this moment.

\appendix
\section{Proof of Lemma~\ref{lem:pro}}\label{app:lem}
\mbox\indent \ref{lem:pro1}:
Let us consider Step 2.1 in the $l$-th loop.
Assuming that $a_l$ is chosen at Step 2.1, $a_l$ satisfies the conditions of Step 2.1. 
That is, $a_l$ is represented as $a_l = t_l^m b_l$ for some $m$, where 
$t_l$ is the periodic string of $a_l$, $t_l \notin T_{l-1}$, and $|b_l|<|t_l|$.

Now we show $m=1$ by contradiction. Suppose that $m \geq 2$.
By the definition of the periodic string, $b_l$ is a prefix of $t_l$.
That is, $b_l$ is represented as $t_l = b_l y$ for some $y$.
Thus $t_l b_l$ is a prefix of $a_l$. Clearly, $t_l b_l$ is also a suffix of $a_l$.
Then $t_l b_l$ is a presuffix of $z_l$, 
because $a_l$ satisfies the condition (i) of Step 2.1.
Since $a_l$ satisfies the condition (ii) of Step 2.1, $t_l$ is not in $T_{l-1}$. 
Then $t_lb_l$, which is not $a_l$ because $m\geq 2$, 
is the shortest string satisfying the conditions of Step 2.1.
This contradicts the fact that $a_l$ is the shortest one. 
Therefore we have $m=1$, that is, $a_l = t_l b_l$. 
\qed

\ref{lem:pro2}: By the property~\ref{lem:pro1}, $a_l = t_l b_l$.
At Step 2.2, we extend $z_{l}$ to the left by $t_{l}^i$. 
Since $b_l$ is a prefix of $t_l$, $t_{l}^i a_{l}$ has $a_{l}$ as a prefix.
Thus $z'_l := t_{l}^i z_{l}$ also has $a_{l}$ as a prefix.

Next, we show that $z_{l+1}$ has $z'_l$ as a prefix.
There are two cases to determine $z_{l+1}$: 
(1) $z_{l+1} := z'_l$ (at Step 2.5);  and (2) $z_{l+1} := u^k z'_l$ (at Step 2.6).
In case of (1), it is obvious that $z_{l+1}$ has $z'_l$ as a prefix.
In case of (2), since we go to Step 2.6, $h \geq j$.
Since $z'_l = t_l^h a_l w$, $z'_l$ has a prefix of $t_l^j a_l$.
Then $uz'_l$ has $u t_l^j a_l = z'_l$ as a prefix, that is, $uz'_l = z'_l w'$ for some $w'$. 
This implies that $u^k z'_l = z'_l (w')^{k}$ and hence $u^k z'_l$ has $z'_l$ as a prefix.
Therefore, $z_{l+1}$ has $z'_l$ as a prefix. 
Since $z'_l$ has $a_{l}$ as a prefix, $z_{l+1}$ also has $a_l$ as a prefix.
\qed

\ref{lem:pro3}: 
Because we extend $z_{l}$ to the left, $z_{l+1}$ clearly has $z_{l}$ as a suffix.
Then $a_{l}$ is also a suffix of $z_{l+1}$. \qed

\ref{lem:pro4}: 
First we show that $a_{l}$ is a presuffix of $a_{l+1}$.
Note that $a_{l}$ is a presuffix of $a_{l+1}$ or $a_{l+1}$ is a presuffix of $a_{l}$,
since by the conditions of Step 2.1 and the property~\ref{lem:pro2}, 
both $a_l$ and $a_{l+1}$ are presuffixes of $z_{l+1}$.
Thus it suffices to assume that $|a_{l+1}| < |a_l|$ and lead to a contradiction.  
Since $z_{l+1}$ has $a_l$ as a presuffix by the properties~\ref{lem:pro2} and~\ref{lem:pro3}, 
$a_{l+1}$ is a presuffix of ${a_l}$.
This implies that $a_{l+1}$ is a presuffix of $z_l$.
Then $a_{l+1}$ satisfies the conditions of Step 2.1 during $l$-th loop.
This contradicts the fact that $a_l$ is such the shortest string.

Second we show that $|a_{l+1}|\neq |a_{l}|$ by contradiction, which implies $|a_{l+1}| > |a_l|$.
Assuming $|a_{l+1}| = |a_l|$, $a_{l+1}=a_{l}$ by the property \ref{lem:pro2}.
Then $t_{l+1}=t_{l}$. However, $t_{l+1}$ is not in $T_l$ by the condition (ii) at Step 2.1, 
which also leads to a contradiction.
\qed

\ref{lem:pro5}: By the property~\ref{lem:pro4}, $a_l$ is a presuffix of $a_{l+1}$ 
and $|a_{l+1}| > |a_l|$. Thus there are some $y$ and $y'$ with $|y| = |y'|>0$
such that $a_{l+1} = a_l y = y' a_l$. Now it suffices to show $|y'| \geq |t_l|$. 
Assuming $|y'| < |t_l|$, the length of the periodic string of $a_l$ is at most $|y'|$.
This contradicts the fact that $t_l$ is the periodic string of $a_l$.
\qed

\ref{lem:pro6}: By contradiction. First we assume $|t_{l+1}| < |t_l|$. 
Noting that $a_l$ is a prefix of $a_{l+1}$ by the property~\ref{lem:pro4}, 
$a_l$ is represented as $a_l = t_{l+1}^m b$ for some integer $m$ and string $b$.
Then the length of the periodic string of $a_l$ should be at most $|t_{l+1}|$. 
This contradicts the fact that $t_l$ is the periodic string of $a_l$. 
Second, assume that $|t_{l+1}| = |t_l|$. This means that 
$t_l = t_{l+1}$ while $t_{l+1}$ is not in $T_l$ by the condition (ii) at Step 2.1, 
which is also a contradiction. \qed

\ref{lem:pro7}: We need to consider the following two cases at Step 2.6: 
(1) $h > j$ and (2) $h = j$.
In case of (1), $|z'_{l}| = h|t_l|+|a_l|+|w| = |u|+j|t_l|+|a_l|$.
Thus $|u| - |t_l| = (h-j-1)|t_l| + |w| > 0$. 
In case of (2), $z'_{l} = u t^j a_l = t^j a_l w$.
Now we show that $|u| > |t_l|$ by contradiction. Assuming that $|u|=|t_l|$, we have $u=t_l$
since $z'_l$ has $t_l$ as a prefix. 
This implies that $z'_l=t^{j+1}_l a_l$, 
which contradicts the definition of $j$ at Step 2.3.
Assuming that $|u|<|t_l|$, the length of the periodic string of $z'_l$ is at most $|u|$.
This implies that the length of the periodic string of $a_l$ is also at most $|u|$, which contradicts
that $t_l$ is the periodic string of $a_l$.
\qed

\ref{lem:pro8}: By the properties \ref{lem:pro5} and \ref{lem:pro6},
$N \geq |a_{l_{max}-1}| \geq \sum_{l=1}^{l_{max}-2} |t_l| \geq (l_{max}-2)^2/2$. 
This implies $l_{max}=O(\sqrt{N})$. \qed

\section{Proof of Theorem~\ref{thm:lower}}\label{app:low}

Our proof consists of two steps.
First, we introduce another oracle model (the {\it AS-oracle}) similar to the S-oracle and
show a lower bound for the query complexity of identifying an AS-oracle.
Secondly, we reduce the identification problem for an AS-oracle to the identification problem for an S-oracle, with some overhead.

To show the lower bound for AS-oracles, we now revisit one version of the (nonnegative) quantum adversary method, 
called {\it the strong weighted adversary method} in \cite{SS06}, due to Zhang \cite{Zha04}.
Let $f$ be a function from a finite set $S$ to another finite set $S'$. 
The goal is to compute $f(x)$, where $x\in S$ is the input. 
In the query complexity model, the input $x$ is given as an oracle. 
More precisely, suppose that the oracle $O_x$ corresponding to $x$ is the unitary transformation
$O_x|q,a,z\rangle=|q,a\oplus\zeta(x;q),z\rangle$,
where $|q\rangle$ is the register for a query string $q$ from a finite set $Q$, 
$|a\rangle$ is the register for the binary answer $\zeta(x;q)$ and $|z\rangle$ is the work register. 
Here $\zeta$ is some function from $S\times Q$ to $\{0,1\}$.
Then the strong adversary method is restated as follows:

\begin{lemma}\label{lem:adv}
Let $w,w'$ denote a weight scheme as follows:
\begin{enumerate}
\item Every pair $(x,y)\in S\times S$ is assigned 
a nonnegative weight $w(x,y)=w(y,x)$ that satisfies $w(x,y)=0$ whenever $f(x)=f(y)$.

\item Every triple $(x,y,q)\in S\times S\times Q$ 
is assigned a nonnegative weight $w'(x,y,q)$ that satisfies $w'(x,y,q)=0$ 
whenever $\zeta(x;q)=\zeta(y;q)$ or $f(x)=f(y)$, and $w'(x,y,q)w'(y,x,q)\geq w^2(x,y)$ 
for all $x,y,q$ such that $\zeta(x;q)\neq \zeta(y;q)$ and $f(x)\neq f(y)$.
\end{enumerate}
For all $x,q$, let $\mu(x)=\sum_y w(x,y)$ and $\nu(x,q)=\sum_y w'(x,y,q)$. 
Then, the quantum query complexity of $f$ is at least
\[
\Omega\left(
\max_{w,w'}\min_{
{\scriptsize \begin{array}{c} x,y,q,\ w(x,y)>0,\\ \zeta(x;q)\neq \zeta(y;q)\end{array} }}
\sqrt{\frac{\mu(x)\mu(y)}{\nu(x,q)\nu(y,q)}}
\right).
\]
\end{lemma}

\subsection{New Oracle Model - Anchored Substring Oracle}
To prove Theorem~\ref{thm:lower}, we introduce another oracle model similar to the S-oracle.
We call it an {\it anchored substring oracle} or, simply, an {\it AS-oracle}.\\
\begin{definition}
An AS-oracle is a binary string $X=(x_0,x_1,\dots ,x_{N-1})$.
A query to the AS-oracle is a pair of an index and a string
$q=(i,s) \in \{0,1,\ldots,N-1\} \times \{0,1\}^*$.
The answer from the AS-oracle is the binary value $\tau(X;q)$ defined as follows:
If the substring $x_ix_{i+1} \cdots x_{i+|s|-1}$ of $X$ is equal to $s$ then $\tau(X;q)=1$,
otherwise $\tau(X;q)=0$.
\end{definition}
We give the following lower bound. 
\begin{lemma}\label{lem:as}
The quantum query complexity for identifying an AS-oracle is $\Omega\left( \frac{N}{\log N} \right)$.
\end{lemma}

{\it Proof.}
Let us assume $N$ is divisible by $\lceil \log N \rceil$ (generalization is easy and omitted). 
We regard the oracle string $X \in \{0,1\}^N$ as 
$X=(X_1,X_2, \ldots , X_{ \frac{N}{ \lceil \log N \rceil }} ) \in [M]^{\frac{N}{\lceil \log N \rceil}}$,
where $X_j \in \{0,1\}^{\lceil \log N \rceil}$ for $j=1, \cdots ,\frac{N}{\lceil \log N \rceil}$
and $M= 2^{\lceil \log N \rceil}$.
For any instances $X,Y$, we define $D(X,Y):=| \{ i | X_i \neq Y_i \} |$ and 
call this quantity the {\it block distance} between $X$ and $Y$.

We now define sets $Q(a,l,b)$ of queries to the AS-oracle 
for any $a,b\in\{0,\ldots,\lceil \log N \rceil -1 \}$ and any $l\in\{0,\ldots,N / \lceil \log N \rceil\}$
such that $(a,l)\neq (0,0)$.
If $a\neq 0$ we define
$$Q(a,0,0)=\{ q=(i,s) \mid  |s|=a \text{ and }
j \cdot \lceil \log N \rceil \leq i < a + i < (j+1) \cdot \lceil \log N \rceil \text{ for some }j\}.$$
A query $q \in Q(a,0,0)$ reads $a$ letters in the same block.
If $l>0$ or $ab>0$, then we define
$$Q(a,l,b)
=\{ q=(i,s) \mid  |s|=a+l\cdot\lceil\log N\rceil +b \text{ and }
a + i \equiv 0\pmod{\lceil \log N \rceil }.$$
A query $q \in Q(a,l,b)$ reads, for some index $j$, 
the last $a$ letters in block $X_{j-1}$, all the letters in the 
$l$ blocks $X_{j}$ to $X_{j+l-1}$, and the
first $b$ letters in block $X_{j+l}$.

To use Lemma~\ref{lem:adv}, let $S=\{0,1\}^N$, $Q = \bigcup_{a,l,b}Q(a,l,b)$, $\zeta(X;q)=\tau(X;q)$, and $f(X)=X$. 
Now we give a weight scheme.
For any pair $(X,Y) \in S\times S$, let $w(X,Y)=1$ 
if $D(X,Y)=1$ and $w(X,Y)= 0$ otherwise.
For any $X, Y$ and $q \in Q(a,l,b)$,
we set the weight $w'(X,Y,q)$ as follows.
If $D(X,Y) \neq 1$ or $\tau(X;q) = \tau(Y;q)$, we set $w'(X,Y,q)=0$.
Otherwise:
\begin{itemize}
\item[1.]
If the index $j$ such that $X_j \neq Y_j$ represents one of 
the $l$ blocks covered in the part read by query $q$, then $w'(X,Y,q)=1$.
\item[2.]
If the index $j$ such that $X_j \neq Y_j$ 
represents a block in which query $q$ reads $a$ letters or $b$ letters, then $w'(X,Y,q)=l+1$ 
when $\tau(X;q)=1$ (and then $\tau(Y;q)=0$) and $w'(X,Y,q)=\frac{1}{l+1}$ when $\tau(X;q)=0$.
\end{itemize}
It is easy to check that this satisfies the conditions of a weight scheme. 
Then, for any $X$, we have
\[
\mu(X)=\sum_Yw(X,Y)=\frac{N}{\lceil \log N \rceil} \cdot (M-1)=\frac{N(M-1)}{\lceil \log N \rceil} = \Omega \left( \frac{N^2}{\log N} \right),
\]
by $M=2^{\lceil \log N \rceil} =\Theta(N)$.
We need to evaluate $\nu(X,q)\nu(Y,q)$ 
for pairs $(X,Y)$ such that $\tau(X;q)=1$ and $\tau(Y;q)=0$, or 
$\tau(X;q)=0$ and $\tau(Y;q)=1$. 
By symmetry, we only consider the case where $\tau(X,q)=1$ and $\tau(Y,q)=0$. Then,
\[
\nu (X,q) = (M-\frac{M}{2^a}) \cdot (l+1) + l \cdot (M-1) +(M-\frac{N}{2^b}) \cdot (l+1) < 3(l+1)M.
\]
The quantity $\nu (Y,q)$ is 1 or $\frac{M}{2^a \cdot (l+1)}$ or $\frac{M}{2^b \cdot (l+1)}$.
In all of these three cases, it satisfies the inequality
\[
\nu (Y,q)  < \frac{M}{l+1}  \:\:\: ( \textrm{since }l \leq N / \lceil \log N \rceil < M).
\]
Hence, 
$\nu (X,q) \nu (Y,q) < 3M^2 =O(N^2)$.

By Lemma~\ref{lem:adv} the quantum query complexity of identifying an AS-oracle is at least
\[
\Omega\left(
\min_{
{\scriptsize \begin{array}{c} X,Y,q,\ w(X,Y)>0,\\ \tau(X;q)\neq \tau(Y;q)\end{array} }}
\sqrt{\frac{\mu(X)\mu(Y)}{\nu(X,q)\nu(Y,q)}}
\right)
=
\Omega\left(\frac{N}{\log N}\right).\hspace{35mm}\qed
\]
\subsection{Reduction}\label{app:red}

We prove the lower bound for identifying an S-oracle (Theorem \ref{thm:lower}) by a reduction 
from the problem of identifying an AS-oracle.

We show how to embed an AS-oracle string of size $N$ into an S-oracle string.
For any AS-oracle string $X=x_0 x_1 \cdots x_{N-2} x_{N-1}$, 
we construct the following S-oracle string $X' \in \{0,1\}^{O(N\log N)}$:
\[
X'=B(0) B(0)^R \sharp x_0 \sharp\sharp B(1) B(1)^R \sharp x_1 \sharp\sharp 
\cdots B(N-1) B(N-1)^R \sharp x_{N-1},
\]
where $B(i)$ is the binary representation of index $i$, $B(i)^R$ is the reverse string of $B(i)$, 
and $\sharp = 1^{10 \log N}$.

First, we can easily see that a query to the AS-oracle string $X$ 
is embedded into a query to the S-oracle string $X'$: 
For a query $(i,z_1\cdots z_m)$ to $X$, the corresponding query to $X'$ is
\[
B(i)B(i)^R\sharp z_{1}\sharp\sharp B(i+1)B(i+1)^R \sharp z_{2} \sharp\sharp 
\cdots B(i+m-1) B(i+m-1)^R \sharp z_{m}.
\] 

Second, we show that any query $s$ to the S-oracle string $X'$ is useless (i.e., 
the answer is independent of $X'$) or corresponds to a query to the AS-oracle $X$. 
Assume that $s$ is not useless. One can consider the following two cases.
\begin{enumerate}
\item The string $s$ is ``long'': 
In this case, the query string includes $B(i)$ or $B(i)^R$, and hence 
which part of $X'$ is referred by the query is determined.
Thus the query string to the S-oracle corresponds to a query string to the AS-oracle. 
For example, if 
\[
s=B(1)^R\sharp z_1\sharp\sharp B(2)B(2)^R\sharp z_2\sharp\sharp B(3),
\]
this query corresponds to the query string $(1,z_1z_2)$ to the $AS$-oracle.

\item The string $s$ is ``short'':
In this case, we cannot determine which part of $X'$ is referred uniquely 
because $s$ corresponds to only a part of $B(i)$ or $B(i)^R$. 
But even for such a case, we still obtain the higher bits of $B(i)$'s 
by our construction of $X'$. 
For example, let us consider a query string $00 \sharp 1$.
Note that the two bits $00$ of this query corresponds to one of the indexes $i$ 
of the query of the AS-oracle such that the highest two bits of $i$ is $00$. 
Then, the query string $00 \sharp 1$ indicates whether at least one bit from $x_{0}$ to $x_{N/4-1}$ 
is $1$ or not. Thus, this query corresponds to the query $(0,0^{N/4})$ of the AS-oracle. 
\end{enumerate}

By the above arguments, we can reduce identifying the AS-oracle to identifying the S-oracle 
with a $O(\log N)$ factor. By Lemma~\ref{lem:as}, the lower bound for the S-oracle 
is then $\Omega\left(\frac{N}{\log ^2 N}\right)$, which completes the proof of Theorem~\ref{thm:lower}.

\section{Proof of Theorem~\ref{thm:res}}\label{app:res}

We prove by using the adversary method.
We define that $Q_{\geq L}$ (resp. $Q_{\leq L}$) is 
the set of query strings $q$ such that $|q| \geq L$ (resp. $|q| \leq L$).
Let $L_1 := 3 \log N$ and $L_2 := \log N - 1 - 2 \log \log N$.
Our weight scheme is as follows:
For any pair $(x,y) \in S \times S$ such that $x\neq y$, let $w(x,y)=1$.
For any triple $(x,y,q) \in S \times S \times (Q_{\geq L_1} \cup Q_{\leq L_2} )$ 
such that $\chi(x;q) \neq \chi(y;q)$, let $w'(x,y,q)=1$.

It is easy to check that this satisfies the conditions of a weight scheme.
Then, for any $x$, we have $\mu(x) = \sum_{y}w(x,y) = 2^N -1$.
For evaluating $\nu(x,q)\nu(y,q)$, 
we only consider the case where $\chi(x;q)=1$ and $\chi(y;q)=0$ by symmetry.
Now $\nu(x,q)$ (resp. $\nu(y,q)$) means the number of instances $y$ (resp. $x$)
such that $\chi(y;q)=0$ (resp. $\chi(x;q)=1$).

When $|q| \geq L_1$,
it holds that $\nu(x,q) \leq 2^N$ and
$\nu(y,q) \leq (N-|q|+1) \cdot 2^{(N-|q|)} < N \cdot 2^{(N-L_1)}= \frac{2^N}{N^2}.$
The value of $\nu(y,q)$ is obtained by considering the case that 
$x_i \cdots x_{i+|q|-1}$ is equal to $q$ for some $i$.
When $|q| \leq L_2$,
it holds that $\nu(x,q) < (2^{|q|}- 1)^{\frac{N}{|q|}} = 2^N \cdot (1-\frac{1}{2^{|q|}})^{\frac{N}{|q|}}
 < 2^N \cdot \left(\frac{1}{2} \right)^{\frac{N}{2^{|q|}|q|}}$,
and $\nu(y,q) \leq 2^N$.
The value $\nu(x,q)$ is obtained by dividing the instance into $N/|q|$ blocks of length $|q|$ 
and considering that none of the blocks are equal to $q$.
Evaluating the value $\left(\frac{1}{2} \right)^{\frac{N}{2^{|q|}|q|}}$,
\[
 \left(\frac{1}{2} \right)^{\frac{N}{2^{|q|}|q|}}
 \leq \left(\frac{1}{2} \right)^{\frac{N}{2^{L_2}L_2}}
 < \left(\frac{1}{2} \right)^{\frac{N \cdot 2\log^2 N}{N \log N }}
 = \frac{1}{N^2}.
\]
Hence for all $q \in Q_{\geq L_1} \cup Q_{\leq L_2}$,
$\nu(x,q)\nu(y,q) \leq \frac{2^{2N}}{N^2}$.
By Lemma~\ref{lem:adv} the quantum query complexity is at least
\[
\Omega\left(
\min_{
{\scriptsize \begin{array}{c} x,y,q, w(x,y)>0,\\ \chi(x;q)\neq \chi(y;q),\\ |q|\geq L_1 \text{ and } |q| \leq L_2 \end{array} }}
\sqrt{\frac{\mu(x)\mu(y)}{\nu(x,q)\nu(y,q)}}
\right)
=
\Omega(N).
\]
\end{document}